 \documentstyle{article}

\let\a=\alpha \let\b=\beta \let\g=\gamma \let\d=\delta \let\e=\epsilon
 \let\h=\eta \let\q=\theta  \let\k=\kappa
\let\l=\lambda \let\m=\mu \let\n=\nu   \let\r=\rho
\let\s=\sigma \let\t=\tau  \let\ve=\varepsilon
  \let\c=\chi \let\y=\psi
\let\w=\omega       \let\D=\Delta  \let\L=\Lambda

\let\la=\label \let\ci=\cite 
  
 \def\bd{\begin{document}} \def\ed{\end{document}}
\def\ds{\documentstyle} \let\fr=\frac \let\bl=\bigl \let\br=\bigr
\let\Br=\Bigr \let\Bl=\Bigl
\let\bm=\bibitem
\let\na=\nabla
\let\pa=\partial \let\ov=\overline
\newcommand{\be}{\begin{equation}}
\newcommand{\ee}{\end{equation}}
\def\ba{\begin{array}}
\def\ea{\end{array}}
\newcommand{\ho}[1]{$\, ^{#1}$}
\newcommand{\hoch}[1]{$\, ^{#1}$}
\newcommand{\bea}{\begin{eqnarray}}
\newcommand{\eea}{\end{eqnarray}}
\newcommand{\ra}{\rightarrow}
\newcommand{\lra}{\longrightarrow}
\newcommand{\Lra}{\Leftrightarrow}
\newcommand{\ap}{\alpha^\prime}
\newcommand{\bp}{\beta^\prime}
\newcommand{\tr}{{\rm tr} }
\newcommand{\Tr}{{\rm Tr} }
\newcommand{\NP}{Nucl. Phys. }
\newcommand{\tamphys}{\it Center for Theoretical Physics\\
Physics Department \\ Texas A \& M University
\\ College Station, Texas 77843}

\begin{document}

\hfill{CTP-TAMU-13/93 }

\vspace{24pt}

\begin{center}

{ \large {\bf BRS Cohomology of the Supertranslations in D=4}}

\vspace{36pt}

J. A. Dixon and R. Minasian

\vspace{6pt}

{\tamphys}

\vspace{6pt}

\vspace{6pt}

\underline{ABSTRACT}

\end{center}
Supersymmetry transformations are a kind of square root of spacetime
translations. The corresponding Lie superalgebra always contains the
supertranslation operator $ \d = c^{\a} \s^{\m}_{\a \dot \b} {\ov c}^{\dot \b}
(\e^{\m})^{\dag} $.  We find that the cohomology
of this operator depends on a spin-orbit coupling in an $SU(2)$ group
and has a quite complicated structure. This spin-orbit type coupling
will turn out to be basic in the cohomology of supersymmetric
field theories in general.

\vfill

\baselineskip=12pt

\pagebreak

\setcounter{page}{1}

\hsize=6.5truein
\hoffset=-0.75truein
\section{Introduction}
\setcounter{equation}{0}
\renewcommand{\theequation}{1.\arabic{equation}}

It is well known that anomalies in gauge theories can be classified as elements
of the cohomology space of the BRS operator acting on the space of
local polynomials \ci{brs,bcr}. Due to locality, all cohomology calculations
are
brought to an algebraic form, and so all details of the space-time topology may
be neglected (see, e.g. \ci{dtv}). Here we are interested in
cohomology classes built by space-time independent `ghosts', so the requirement
of locality is trivial, but we shall see many common features
between our construction and cohomology classes of local polynomials.
In particular, the present computation is similar to the analysis of the
cohomology
of supersymmetric theories which was commenced in \ci{cmp2}.

We think the method  describing the BRS cohomology of the supertranslation
operator in four dimensions  outlined in this paper  is directly
applicable to problems which lie at the heart of all four dimensional
supersymmetric quantum field theories and  superstring
theory.  There also may be a simple connection between the present
results and the results in quantum field theory, the most recent discussion
of which one may find in \ci{dhtv}.

Since the supertranslation operator analysed here has a direct analogue
for gauge theories, let us recall an essential point in
the problem of the computation of the BRS cohomology of Yang-Mills
theory before  presenting our calculations.
The golden nugget
in that problem  \ci{brs,cmp1}
is the cohomology of the following nilpotent operator
\be
\d = f^{abc} \w^a \w^b (\w^c)^{\dag}
\la{ym}
\ee
where
\be
\{ (\w^a)^{\dag},
\w^b \} = \d_{a}^{b}
\ee
$(\w^a)^{\dag}$ is an annihilation operator
and $f^{abc}$ are the structure constants of a compact semisimple
Lie algebra.
The cohomology space of this operator
is well known and  is given by the kernel of the  `Laplacian':
\be
\D = (\d + \d^{\dag}) = Y^a Y^a
\ee
where
\be
Y^a = (Y^a)^{\dag} = i
f^{abc} \w^b (\w^c)^{\dag}
\ee
is the generator of rotations.  The kernel consists of all solutions
of
\be
Y^a H = 0
\ee
which are just the invariant antisymmetric tensors of the
adjoint representation for the group.

By a straightforward mapping, these antisymmetric tensors
in turn generate the cohomology space of Yang-Mills theory, which
determines all possible anomalies of such theories.
Gravity works in an analogous way--the relevant group there is
SO(1,D-1) for D dimensional gravity.

The supertranslation operator which  will be  analysed in
the present paper is:
\be
\d = c^{\a} \s^{\m}_{\a \dot \b} {\ov c}^{\dot \b} (\e^{\m})^{\dag}
\la{st}
\ee
As  mentioned, (\ref{st}) is the analogue of (\ref{ym}) for supersymmetry.
Similar operators occur in all supersymmetry theories in all dimensions,
however, their cohomology is certainly dimension dependent.  Here we
concentrate
on the $D=4$ case, and it turns out that performing calculations in  tems of
two component Weyl spinors rather than in terms of Majorana spinors simplifies
the problem significantly. A set of conventions which are particularly well
suited to our problem is introduced in appendix A. They preserve
the symmetry of the theory under complex
conjugation. This symmetry is obscure in most sets of
conventions.  We also give a number of identities  in this notation
for convenience and ease of the present and future work.

The easiest way to see (\ref{st}) arising is to consider the algebra of the
well known supersymmetry translation operators $Q$:
\be
Q_{\a} = \fr{\pa}{\pa \q^{\a} } + \fr{1}{2}
\s^{\m}_{\a \dot \b} {\ov \q}^{\dot \b} \pa_{\m}
\ee
\be
[ Q_{\a}]^* =
{\ov Q}_{\dot \a} =
\fr{\pa}{\pa {\ov \q}^{\dot \a} } + \fr{1}{2}
{\ov \s}^{\m}_{\dot \a \b} {\q}^{ \b} \pa_{\m}
\ee
To close the algebra of these operators we must add an anticommuting
translation `ghost' $\e^{\m}$ and a commuting Weyl spinor ghost
$c_{\a}$ and its complex conjugate ${\ov c}_{\dot \a}$.
The resulting nilpotent `BRS' type operator
is:
\be
\d = c^{\a} Q_{\a}
+
{\ov c}^{\dot \a} {\ov Q}_{\dot \a}
-
c^{\a} \s^{\m}_{\a \dot \b} {\ov c}^{\dot \b}
\e^{\m \dag}
+
\e^{\m}
\pa_{\m}
\ee
which summarizes the algebra of the Q's.
Note that $\d = {\ov \d} $ is a `real' anticommuting operator.
The following identities hold:
\be
\{ Q_{\a},  {\ov Q}_{\dot \b} \}
=
 \s^{\m}_{\a \dot \b}
\pa_{\m}
\ee
Now in a field theory, the operators $Q_{\a}$ and $\pa_{\m}$ become
functional derivative operators whose action is really
quite complex and the cohomology of the operator $\d$ becomes
also quite complicated.  However, the supertranslation part of
this operator remains even in field theory, and
using the present results and the method of
spectral sequences the full BRS cohomology can now be found for chiral
superfields in $D=4$,
$N=1$ supersymmetry \ci{dm1}, although there are still
unsolved difficulties when vector superfields
(i.e. super Yang-Mills fields) are present.

\section{Laplacian}
\setcounter{equation}{0}
\renewcommand{\theequation}{2.\arabic{equation}}

The  calculation of the cohomology space is based on the simple result:
\be
H(\d)
= {\rm ker}\; \d / {\rm im} \; \d
\approx [ {\rm ker}\; \d]^{\perp} \cap [ {\rm im} \; \d ]^{\perp}
= {\rm ker}\; \D
\ee
where $H(\d)$ is the cohomology space of $\d$ and
$\D$ is the Laplacian formed from $\d$:
\be
\D = [ \d + \d^{\dag}]^2
= \d \d^{\dag}
+ \d^{\dag} \d
\ee
Though in this paper we identify $H$ and ${\rm ker}\; \D$ ,in general,
it is important to keep in mind that the
former is actually a factor space which is isomorphic with the latter.

Using various identities contained in appendix A, we show that $\d$ is
an operator in a Fock space with positive definite
metric.  An essential point is that
this can be done even though our spacetime
is intrinsically Minkowskian. The detailed discussion of this point
is presented in appendix B.
After some algebra, we find that the Hodge `Laplacian' operator for the
operator
(\ref{st}) can be written in the form:
\be
\D =
[n + {\ov n} + 2] N +  2  n  {\ov n}
+ 4 [ J_i L_i + {\ov J}_i {\ov L}_i]
\la{lap}
\ee
where we use the abbreviations:
\be
n = c^{\a} (c^{\a})^{\dag}
\la{countc}
\ee
\be
{\ov n} = {\ov c}^{\dot \a} ({\ov c}^{\dot \a})^{\dag}
\ee
\be
N = \e^{\m} (\e^{\m})^{\dag}
\ee
\be
J_i  = \fr{1}{2}
c^{\a} (\s_i)_{\a}^{\;\;\;\b} (c^{\b})^{\dag}
\ee
\be
{\ov J}_i  = \fr{1}{2}
{\ov c}^{\dot \a} ({\ov \s}_i)_{\dot \a}^{\;\;
\dot \b} ({\ov c}^{\dot \b})^{\dag}
\ee
\be
L_i (\s_i)_{\a}^{\;\;\;\b}
=
\fr{1}{2}
(\s^{\m \n})_{\a}^{\;\;\;\b}
\e_{\m} (\e^{\n})^{\dag}
\ee
The operator $L_i$ can also be written more explicitly in the form:
\be
L_i = -\fr{1}{2}
(\e_0 (\e_i)^{\dag}
+ \e_i (\e_0)^{\dag}
+ i \ve_{ijk} \e_j (\e_k)^{\dag})
\ee

It is easy to verify that $L_i$ and $J_i$ obey the
commutation rules of the $SU(2)$ Lie algebra:
\be
[J_i, J_j] = i \ve^{ijk} J_k
\ee
\be
[L_i, L_j] = i \ve^{ijk} L_k
\ee
The complex conjugate equations are:
\be
[{\ov J}_i, {\ov J}_j] = - i \ve^{ijk} {\ov J}_k
\ee
and we note
\be
[J_i, L_j] = 0
\ee

Initially, it was a surprise for us to find that this operator involves only
angular momentum in the compact $SU(2)$ algebra even though
the supertranslation operator is intrinsically defined
in Minkowski space.  But the result is reasonable.
This happens  because
the Fock space is positive definite, so that any relevant
group theory for our Laplacian must necessarily also
be defined in a compact positive definite context.  Anyway,
regardless of the philosophy,
one finds the above result.  The same kind of thing
happens for the BRS operators of string theory when the problem
is formulated in this way.

\section{Discussion of the Laplacian}
\setcounter{equation}{0}
\renewcommand{\theequation}{3.\arabic{equation}}

Since the Laplacian $\D$ in (\ref{lap}) consists solely of
counting operators and coupled angular momentum operators,
finding its kernel is an exercise in the theory of angular momentum.

It is well known that the eigenvalues of angular
momentum operators $J^2$ are of the form
$j (j+1)$ where $j = 0, \fr{1}{2}, 1,
\fr{3}{2}, 2 \ldots$.  We rewrite the above Laplacian in the form:
\be
\D = [n + {\ov n} + 2] N +  2  n  {\ov n}
+ 2 [ K^2 - J^2 - L^2 ]
+ 2 [ {\ov K}^2 - {\ov J}^2 - {\ov L}^2 ]
\ee
where we define the composite angular momentum operators
\be
K_i = J_i + L_i
\ee
\be
{\ov K}_i = {\ov J}_i + {\ov L}_i
\ee
\be
J^2 = J_i J_i ; \;\;\;\;\; L^2 = L_i L_i
\ee
What do the eigenvectors of these various angular momentum
operators look like?

The eigenvectors of $J^2$ are:
\be
 J^2  c^{\a_1} c^{\a_2} \cdots  c^{\a_m} = \fr{m}{2}( \fr{m}{2} +1)
 c^{\a_1} c^{\a_2} \cdots  c^{\a_m}
\;\;\;
m \geq 1
\ee
These expressions are automatically symmetric under interchange
of any pair of indices.

The eigenvectors of $L^2$ are best described using the variables
\be
\e_{\a \dot \b} =
\s^{\m}_{\a \dot \b} \e_{\m}
\la{epsi}
\ee
The $L^2$ eigenvalue of a product of these variables is equal
to the number of symmetrized undotted indices in the product.
Since $\e$ anticommutes, there are a very limited number of
possibilities that are nonzero:
\be
L^2 \e_{\m} = \fr{3}{4}  \e_{\m}; \;\;\;\;\; (l=\fr{1}{2})
\ee
\be
L^2 \e_{\m} \e_{\n} \s^{\m \n }_{\a \b}  = 2
 \e_{\m} \e_{\n} \s^{\m \n }_{\a \b}; \;\;\;\;\; (l=1)
\ee
\be
L^2 \e_{\m} \e_{\n} {\ov \s}^{\m \n }_{\dot \a \dot \b} = 0; \;\;\;\;\; (l=0)
\ee
\be
L^2 \e_{\m} \e_{\n} \e_{\l} =
\ve_{\m \n \l \r} L^2 W^{\r}
=
\fr{3}{4}
\ve_{\m \n \l \r} W_{\r}   ;
\;\;\;\;\; (l=\fr{1}{2})
\ee
with
\be
W_{\r}    = - \fr{1}{6}
\ve_{\m \n \l \r} \e^{\m} \e^{\n} \e^{\l}
\ee
where we note that a vector $W^{\a \dot \b}=(\s^ {\r})^{\a \dot \b} W_{\r}$
can only have one undotted spinor index.
Finally,
\be
L^2 \e_{\m} \e_{\n} \e_{\l} \e_{\r} ; \;\;\;\;\; (l=0)
=0
\ee
For the complex conjugate operators the discussion is identical
except that only dotted indices are relevant.

For the combined operator $K^2$ the eigenvectors have spin $k$ where
$2k$ is the number of symmetric (and consequently uncontracted)
free undotted indices in the eigenvector.  These indices may
come from either $c_{\a}$ or $\e_{\a \dot \b}$.
So the eigenvectors of $K^2$ are of the form (for example):
\be
K^2
\e_{\m} \s^{\m }_{\a \dot \b} c^{\a}  = 0; \;\;\;\;\; (k=0)
\ee
\be
K^2
(\e_{\a \dot \b} c_{\b} + \e_{\b \dot \b} c_{\a})=
2(\e_{\a \dot \b} c_{\b} + \e_{\b \dot \b} c_{\a}); \;\;\;\;\; (k=1)
\ee
etc.

As is known from the theory of angular momenta, the eigenvalues of
$K^2$ run from $|j-l|$ to $(j+l)$.
Now, finding all the possible combinations of eigenvalues that
can give zero for the value of the Laplacian above allows the
identification of the cohomology space.
We do this simply by examining the various cases as a function of
$N= N(\e)$ for $N=0,1,2,3,4$.

\begin{enumerate}
\item $N=0$

For this case, the Laplacian reduces simply to
\be
\D = 2 n {\ov n}
\ee
and the zero eigenvectors are those polynomials which are
independent of either $c$ or of $\ov c$ so that the
product $n {\ov n}$ is zero.
\item $N=1$

For this case, $l= \fr{1}{2}$ and
so $k= \fr{n}{2} + \fr{1}{2}$
or  $k= \fr{n}{2} - \fr{1}{2}$, similarly for
$\ov l$ and  $\ov k$.  Hence we get
\be
2 [ K^2 - J^2 - L^2 ] = n+1
; \;\; {\rm when} \;\;
k= \fr{n}{2} + \fr{1}{2}
\ee
\be
2 [ K^2 - J^2 - L^2 ] = - (n+2) ; \;\; {\rm when} \;\;
k= \fr{n}{2} - \fr{1}{2}
\ee
and similarly for the complex conjugates.
There are then four simple cases to analyze.
One of these is the following:
\[
\D = [n + {\ov n} + 2] +  2  n  {\ov n}
\]
\be
-   (n+2) - 2({\ov n} +2) =
2  n  {\ov n}
- 2
; \;\; {\rm when} \;\;
k= \fr{n}{2} - \fr{1}{2}\;\;
{\rm and} \;\;
{\ov k}= \fr{{\ov n}}{2} - \fr{1}{2}
\la{angmom}
\ee
whose unique solution is $n = {\ov n} = 1$.
The only other solution occurs when ${\ov n} =0$ and
$k= \fr{n}{2} - \fr{1}{2}$ (and the complex conjugate of this).
\item $N=2$

Because of the identity (recall definition (\ref {epsi}))
$\e_{\a \dot \b} \e_{\g \dot \d}= \fr{1}{2}
( \e_{\m}  \e_{\n} \s^{\m \n}_{\a \g} \e_{\dot \b \dot \d}+
 \e_{\m}  \e_{\n} \ov{\s}^{\m \n}_{\dot\b \dot \d} \e_{\a \g} ) $,
there are two possibilities:
$l= 0, {\ov l}= 1$ or
$l= 1, {\ov l}= 0$.

First, consider the case where $l= 1, {\ov l}= 0$.
Then  ${\ov k}= \fr{n}{2}$ and $k$ can take three
possible values
$k= \fr{n}{2} +1, \;
\fr{n}{2}, \; {\rm or} \; \fr{n}{2} - 1 $.
Then
\be
2 [ {\ov K}^2 -  {\ov J}^2 - {\ov L}^2 ] = 0
\ee
and
\be
2 [ K^2 - J^2 - L^2 ] = 2n
; \;\; {\rm when} \;\;
k= \fr{n}{2} + 1
\ee
\be
2 [ K^2 - J^2 - L^2 ] = - 4 ; \;\; {\rm when} \;\;
k= \fr{n}{2}
\ee
\be
2 [ K^2 - J^2 - L^2 ] = - 2(n+2)  ; \;\; {\rm when}\;\;
k= \fr{n}{2} -1
\ee
and similarly for the complex conjugates.
For the last case we get:
\[
\D = 2 [n + {\ov n} + 2] +  2  n  {\ov n}
\]
\be
-  2  (n+2)
=
2 (1 + n) {\ov n} = 0
; \;\; {\rm when} \;\;
k= \fr{n}{2} - 1
\ee
whose unique solution is ${\ov n} = 0$.
This is the only case that yields a solution.
The complex conjugate also works, of course.
\item $N=3$

Now we have again $l = {\ov l} = \fr{1}{2}$
There are no solutions
of the equations for this case.
\item $N=4$

Now we have $l = {\ov l} = 0$
Again there are no solutions.
\end{enumerate}

This concludes the analysis, and the results are presented and summarized
in the next section.

\section{The cohomology space}
\setcounter{equation}{0}
\renewcommand{\theequation}{4.\arabic{equation}}

To summarize, we have shown that
the following set of polynomials constitute all the
vectors in the cohomology space of the operator (\ref{st}).
\be
c^{\a_1} c^{\a_2} \cdots  c^{\a_m}
\;\;\;
m \geq 1
\la{cohom1}
\ee
\be  \e_{\m}
\s^{\m}_{\a \dot \b}  c^{ \a}
 c^{ \a_1}  c^{ \a_2}
\cdots
 c^{ \a_m} \; \; m \geq 0
\la{cohom2}
\ee
\be  \e_{\m} \e_{\n}
\s^{\m \n }_{\a \b} {c}^{\a} {c}^{\b}
{ c}^{ \a_1}
{ c}^{ \a_2}
\cdots
{ c}^{ \a_m} \;\;\;
m \geq 0
\la{cohom3}
\ee
and their complex conjugates:
\be
{\ov c}^{\dot \a_1} {\ov c}^{\dot \a_2} \cdots  {\ov c}^{\dot \a_m}
\;
m \geq 1
\ee
\be  \e_{\m}
\ov\s^{\m}_{\dot\a  \b} {\ov c}^{\dot \a}
 {\ov c}^{\dot \a_1}  {\ov c}^{\dot \a_2}
\cdots
 {\ov c}^{\dot \a_m} \; \; m \geq 0
\ee
\be  \e_{\m} \e_{\n}
\ov\s^{\m \n }_{\dot \a \dot \b} {\ov c}^{\dot \a} {\ov c}^{\dot \b}
{\ov c}^{\dot \a_1}
{\ov c}^{\dot \a_2}
\cdots
{\ov c}^{\dot \a_m} \;\;\;
m \geq 0
\ee

Finally, there is one more polynomial in the
cohomology space which is equal to its complex conjugate:
\be \e_{\m}
c^{\a} \s^{\m}_{\a \dot \b} {\ov c}^{\dot \b}
\la{cohom4}
\ee

\section{Projection Operators for Supertranslations}
\setcounter{equation}{0}
\renewcommand{\theequation}{5.\arabic{equation}}

Since our actual interest is in the use of these results
in finding the cohomology space for supersymmetric field
theory where the use of spectral sequences will be necessary,
we have found the orthogonal projection operators which project onto
the space $H$.
These operators all satisfy the relations
\be
\Pi = \Pi^{\dag} = \Pi^2
\ee

The projection operator onto (\ref{cohom1}) has the form
$\Pi_{N=0} \Pi_{\ov n=0} \Pi_{n=r} \; (r \geq 1)$ where
\be
\Pi_{ n  = r } = \sum_{l=r}^{\infty}
\fr{(-1)^{(l-r)}  }{(l-r)!} n_l
\la{pro1}
\ee
and we remind the reader that $n$ is an operator
defined in (\ref{countc}).  The operators $n_l$ are defined
recursively by
\be
n_{l+1} = (n - l) n_l
\ee
and are distinguished by the fact that they are normal ordered operators.
For example,
\be
n_2 = c^{\a} c^{\b}
[c^{\a} c^{\b}]^{\dag}
\ee
This operator satisfies the relation
\be
n \Pi_{ n  = r } \equiv
n_1 \Pi_{ n  = r } =
r \Pi_{ n  = r }  \;\;\;\;\; r=0,1,2, \cdots
\ee

Next, we find the projection operator onto (\ref{cohom4}) $\e_{\m}
c^{\a} \s^{\m}_{\a \dot \b} {\ov c}^{\dot \b}$.
Here we must first take the combination of three operators of type (\ref{pro1})
$\Pi_{N=1} \Pi_{\ov n=1} \Pi_{n=1}$ which selects the subspace
$N= n = {\ov n} =1$. (\ref{cohom4}) is
the totally contracted vector from this subspace.  We find that
the following operator performs the task of picking out the contracted part:
\be
\Pi = ( \fr{1}{4} - J \cdot  L )
( \fr{1}{4} - {\ov J} \cdot  {\ov L} )
\Pi_{N=1} \Pi_{\ov n=1} \Pi_{n=1}
\la{pro2}
\ee
where
\be
J \cdot  L
= J_i L_i
\ee
To show this, one uses the following result  which is easily derived from
(\ref{angmom}):
\be
(J \cdot  L) (J \cdot  L) \Pi_{N=1} \Pi_{\ov n=1} \Pi_{n=1} =
( \fr{3}{16} - \fr{1}{2}  J \cdot  L )
\Pi_{N=1} \Pi_{\ov n=1} \Pi_{n=1}
\ee
This can be used to show that, for the operator (\ref{pro2}), we have
\be
(J \cdot  L)  \Pi = - \fr{3}{4} \Pi
\ee

The third projection operator is the one onto the vector
$ \e_{\m}
\s^{\m}_{\a_1 \dot \b}  c^{ \a_1}
 c^{ \a_2}  c^{ \a_3}
\cdots
 c^{ \a_r}$ , for $r \geq 1$.
As above, first we project onto the space where
$N=1, n=r, {\ov n} =0$.
Then in this subspace we need
\be
J \cdot L \Pi = - [\fr{r}{4}  + \fr{1}{2} ] \Pi
\ee
The operator which accomplishes this is:
\be
\Pi =
 [ \fr{r}{2(r+1)} -\fr{2}{r+1} J \cdot  L ]
\Pi_{N=1} \Pi_{\ov n=o} \Pi_{n=r}
\la{pro3}
\ee

For the vectors $\e_{\m} \e_{\n}
c^{\a} \s^{\m \n}_{\a \b} {c}^{\b} c^{\a_1} \cdots c^{\a_{r-2}}$
for $r \geq 2$,
we must first project onto the relevant subspace
with
$ \Pi_{N=2, n= r , {\ov n}= 0}$,
then onto the subspace in which $L^2 =2$.
This can be accomplished with the operator:
\be
\Pi_{L^2 =2}  = [ \fr{1}{2} + \fr{1}{8} (R + R^{\dag} )]
\ee
where
\be
R =
2 i \e_{ijk} \e_0 \e_i \e_j^{\dag}
\e_k^{\dag}
\ee
To see that this has the desired properties, we note that
\be
L^2 \Pi_{L^2 =2} = [  1 + \fr{1}{4} ( R + R^{\dag}) ]
\Pi_{L^2 =2} = 2 \Pi_{L^2 =2}
\ee
This can be shown using
the results that
\be
R R^{\dag}
\Pi_{ N = 2}
=
16 \e_0 \e_0^{\dag}
\Pi_{ N = 2}
\ee
\be
R^{\dag} R
\Pi_{ N = 2}
=
16 ( 1 - \e_0 \e_0^{\dag})
\Pi_{ N = 2}
\ee
The next step is the projection onto the subspace where
the eigenvalue $k(k+1)$ of $K^2$ is
$ k = \fr{r}{2} -1$, or equivalently,
we need:
\be
J \cdot L \Pi = - [\fr{r}{2} + 1 ] \Pi
\ee
The necessary operator is:
\be
\Pi =  [ \fr{r-1}{4(r+1)} -\fr{r-1}{r(r+1)} J \cdot  L
+\fr{2}{r(r+1)} L_{ij} J_{ij} ]
\Pi_{N=2} \Pi_{\ov n=0} \Pi_{n=r} \Pi_{L^2 =2}
\la{pro4}
\ee
where we define the normal ordered expressions $J_{ij}$
and $L_{ij}$ by:
\be
J_i J_j = \fr{1}{4} n \d_{ij} + \fr{1}{2} i \e_{ijk} J_k + J_{ij}
\ee
\be
L_i L_j = \fr{1}{4} N \d_{ij} + \fr{1}{2} i \e_{ijk} L_k
+ L_{ij}
\ee
and explicitly
\be
J_{ij} = c^{\a} c^{\b}
(\s_i)_{\a}^ {\,\ \g}
(\s_j)_{\b}^ {\,\ \d} (c^{\g})^{\dag}  (c^{\d})^{\dag}
\ee

The total projection operator for the cohomology space of the supertranslation
operator in $D=4$ is the sum of (\ref{pro1}), (\ref{pro2}), (\ref{pro3}),
(\ref{pro4}) and the complex conjugates (where necessary).

\section{Conclusion}

The Laplacian operator for the supertranslation operator can be written in
a very simple and transparent form  (\ref {lap}) which contains only
counting and angular momentum operators.  Then it is straightforward, though
somewhat involved, to deduce the form of the kernel of this Laplacian and so
the cohomology space.  The cohomology space is described in equations
(\ref{cohom1} - \ref{cohom4}). Its complexity is a reflection of the complexity
of the cohomology of supersymmetric theories in general.

Surprisingly, the cohomology calculation involves only the compact Lie algebra
$SU(2)$ even though supersymmetry is a Minkowski space algebra. This is a
consequence of the positive definite metric on Fock space.
Very similar things happen for the general BRS cohomology
of the chiral superfield in supersymmetric quantum field
theory, and we intend to treat various such theories in
forthcoming papers.

The cohomology space of super Yang-Mills involves
vector superfields as well as chiral superfields, and
we have not yet been able to find the
cohomology of that theory.  The difficulty is that in the
cohomology problem for super
Yang-Mills theory,
the
gauge transformations and the supersymmetry transformations
seem to become entangled in a complicated way, even
though they are rather well separated in the relevant
$\d$ operator.

The BRS cohomology of
supergravity is, of course, a separate
problem and a very complicated one.  But the supertranslation
operator treated here naturally appears in
supergravity and we hope that the present results
will be helpful there.

\appendix
\section{Conventions and Useful Formulae}
\setcounter{equation}{0}
\renewcommand{\theequation}{A.\arabic{equation}}

Though many different sets of conventions exist in literature
(see the useful discussion in \ci{soh}), we do not quite follow them because of
our desire to have symmetry under complex conjugation.
Our Lorentz metric is defined by the relation:
\begin{equation} x_{\mu} x^{\mu} = \eta_{\mu \nu} x^{\mu} x^{\nu}
=  - x_{0}^{2} + x_{1}^{2} + x_{2}^{2} + x_{3}^{2}
\la{a1}
\end{equation}
Lorentz transformations $\L^{\m}_{\; \n}$ are real matrices which
preserve this quadratic form when they are used to transform
the vectors:
\be
x'^{\m} =
\L^{\m}_{\; \n}
x^{\n}
\ee
so they satisfy:
\be
\L^{\m}_{\; \n}
\L^{\s}_{\; \t}
\h_{\m  \s}
=
\h_{\n  \t}
\ee
Another way to preserve this quadratic form is to consider
the following action of
the group SL(2,C) of all of $2 \times 2$ complex matrices $M$
of determinant $1$ on a $2 \times 2$ complex hermitian matrix:
\be \s^{\m} x'_{\m}
=
M \s^{\m} x_{\m} M^{\dag}
\la{msm}
\ee
where $\s^{\m}$ are a basis of the set of
$2 \times 2$ complex hermitian matrices, which are chosen to be Pauli matrices.
The transformation (\ref{msm})
defines a Lorentz transformation on the four vector
\be
x'^{\m} = \L^{\m }_{\; \n} x^{\n}
\ee
since the determinant is preserved by this transformation
and it is equal to (minus) the quadratic form (\ref{a1}):
\be
\det
 [ M \s^{\m} x_{\m} M^{\dag}]
 = det [\s^{\m} x'_{\m}]
=  det [\s^{\m} x_{\m}]
=  x'_0 x'_0
- x'_i x'_i = x_0 x_0
- x_i x_i \ee
We will write the indices of the matrix M in the following
positions:
\be
M_{\a}^{\; \; \b}
\ee
and its complex conjugate has the indices:
\be
(M_{\a}^{\; \; \b})^*
=
{\ov M}_{\dot \a}^{\; \; \dot \b}
\ee
and  its hermitian conjugate has indices:
\be
(M_{\a}^{\;\; \b})^{*T}
=
(M_{\a}^{\; \; \b})^{\dag}
=
({\ov M}^T)^{ \dot \b}_{\; \; \dot \a}
\ee
The notation $A^*$ is interchangeable with the notation
$\ov A$.  Both mean simply complex conjugation.
It follows that the position and kind of indices of the matrices $\s^{\m}$
are now determined to be of the form:
\be
\s^{\m}_{ \a \dot \b} = ( 1,  \s^i )_{\a \dot  \b}
\ee
Since the $\s$ matrices are hermitian,
the complex conjugate matrices are defined by:
\be
(\s^{\m}_{\a \dot \b})^*
=
\ov{\s}^{\m}_{\dot \a \b}
=
\s^{\m}_{\b \dot \a}
\ee

Contrary to the usual convention,
we do not reverse the order of (anticommuting)
spinors when taking the complex conjugate.
Such a change of order spoils the natural symmetry
of supersymmetry under complex conjugation and makes all
computations more tricky. (Reversing the
order of commuting spinors makes no difference of course.)
Indices are raised and lowered as follows:
\be
\y^{\a} = \ve^{\a \b} \y_{\b}
\ee
\be
\y_{\a} =  - \ve_{\a \b} \y^{\b}
\ee
\be
\ov{\y}^{\dot \a} = \ve^{\dot \a \dot \b} \ov{\y}_{\dot \b}
\ee
\be
\ov{\y}_{\dot \a} =  - \ve_{\dot \a \dot \b} \ov{\y}^{\dot \b}
\ee
where the $\ve$ tensors are real antisymmetric matrices with:
\be
\ve^{ \a  \g}
\ve_{ \b  \g}
=
\ve_{\b}^{\;\;\a}
=
-
\ve^{\a}_{\;\;\b}
=
\d^{\a}_{\b}
\ee
\be
\ve^{\dot \a \dot \g}
\ve_{\dot \b \dot \g}
=
\d^{\dot \a}_{\dot \b}
\ee
where  $\d^{\a}_{ \b} = 1 $ if $\a = \b $ and $\d^{\a}_{ \b} = 0 $
if
$\a \neq \b$ (Same for
$\d^{\dot \a}_{\dot \b}$).
We take:
\be
\ve^{\a \b} =
i (\s_2)^{\a \b}
\ee
\be
\ve_{\a \b} =
i (\s_{ 2})_{ \a \b}
\ee
We also have:
\be
\q^{\a} \q^{\b}
= -
\fr{1}{2}
\q^2 \ve^{\a \b}
\ee
\be
\q_{\a} \q^{\b}
= -
\fr{1}{2}
\q^2 \d_{\a}^{ \b}
\ee
where
\be
\q^2 = \q^{\a}
\q_{\a}
\ee
and
\be
\q_{\a} \q_{\b}
= -
\fr{1}{2}
\q^2 \ve_{\a \b}
\ee
Using this rule for raising and lowering indices,
we get:
\be
\ov{\s}^{\m \dot \a \b}
=
\ve^{ \dot \a \dot \d}
\ve^{ \b \g}
\ov{\s}^{ \m}_{\dot  \d \g}
= ( 1, - \s^i )^{\dot \a \b}
\ee
so that:
\be
\ov{\s}_{\m}^{ \dot \a \b}
=
- ( 1,  \s^i )^{\dot \a \b}
\la{-sigma}
\ee
where we use the relations:
\be
\s_2 (\s_i)^* \s_2
= -
\s_i
\ee
The well known relation for Pauli matrices
\be
\s_{i} \s_{j} = \d_{ij} {\bf 1} +  i \ve_{ijk} \s_{k}
\ee
results in a number of other relations such as:
\be
\s^{\m}_{\a \dot{\a}}
\ov{\s}^{\n \dot{\a} \b }
+
\s^{\n}_{\a \dot{\a}}
\ov{\s}^{\m \dot{\a} \b }
=
 - 2 \eta^{\m \n} \d^{\b}_{\a}
\ee
which yields
\be \ve^{\dot{\g} \dot{\d}}
\s^{\m}_{\a \dot{\g}}
\s^{\n}_{\b \dot{\d}}
\pa_{\m} \pa_{\n}
=
 -  \ve_{\a \b}
\Box
\ee
\be
\s^{\m}_{\a \dot{\b}}
\ov{\s}_{\m}^{ \dot{\g} \d }
=
- 2 \d^{\d}_{\a}
 \d^{\dot \g}_{\dot \b}
\ee
We define:
\[
\s^{\m \n}_{\a \b} =
\s^{\m \n}_{\b \a} =
- \s^{\n \m}_{\a \b} =
\fr{1}{2} [
\s^{\m}_{\a \dot{\g}}
\ov{\s}^{ \n \dot{\g}}_{\; \;\;  \; \b } -
\s^{\n}_{\a \dot{\g}}
\ov{\s}^{ \m \dot{\g}}_{\; \; \; \;  \b }]
\]
\be
 =
\fr{1}{2} [
\s^{\m}_{\a \dot{\g}}
\s^{\n}_{\b \dot{\d}}
-
\s^{\n}_{\a \dot{\g}}
\s^{\m}_{\b \dot{\d}}
]
\ve^{\dot{\g} \dot{\d} }
\ee
Then,
\be
(\s^{\m \n})_{\a \b}
=
\s^{\m}_{\a \dot{\b}}
\;
\ov{\s}^{\n \dot{\b}}_{\; \;\;  \g }
+ \eta^{\m \n} \ve_{\a \b}
\ee
\be
\s^{\m}_{\a \dot \b}
{\ov \s}^{\n {\dot \b} \g}
=
( \s^{\m \n})_{ \a}^{\;\;  \g}
- \h^{\m \n} \d_{\a}^{\; \;\g}
\la{sigma}
\ee
\be
\ve_{\m \n}^{ \; \; \; \;  \k \l}
\; \; (\s^{\m \n  })_{\a \b }
= 2i
(\s^{\k \l })_{\a \b }
\ee

\be
\s^{\m\n}_{\a \b}
\s^{\l \b }_{\dot \g }
=
\s^{\m  }_{\a \dot \g } \h^{\n \l}
-
\s^{\n  }_{\a \dot \g } \h^{\m \l}
+i \ve^{\m \n \l \r} \s_{\r  \a \dot \g }
\ee

\[
(\s^{\m \n})_{\a  \b}
(\s^{\l \t})^{\b}_{\; \; \g}
=
2 \h^{\m \l} (\s^{\n \t})_{\a  \g}
+
2 \h^{\n \t} (\s^{\m \l})_{\a  \g}
-
2 \h^{\m \t} (\s^{\n \l})_{\a  \g}
\]
\be
-
2 \h^{\n \l} (\s^{\m \t})_{\a  \g}
+
[
 \h^{\m \l} \h^{\n \t} -
 \h^{\m \t} \h^{\n \l})
+i
\ve^{ \m \n \l \t}
]
 \ve_{\a  \g}
\ee
The same holds for complex conjugates:
\[
\ov{\s}^{\m \n \dot{ \a} \dot {\b}} =
\ov{\s}^{\m \n \dot{ \b} \dot {\a}} =
-\ov{\s}^{\n \m \dot{ \a} \dot {\b}}
\]
\be
= -
\fr{1}{2} [
\ov{\s}^{ \m \dot{\a} \g }
\s^{ \n \d \dot{\b} }
-
\ov{\s}^{ \n \dot{\a} \g }
\s^{ \m \d \dot{\b} }
]\ve_{\g \d }
\ee
Note that $\s^{0i}$ is not independent of $\s^{ij}$:
\be
(\s^{0i})_{\a}^{\; \; \b}
= - (\s^i)_{\a}^{\; \; \b}
\ee
\be
(\s^{ij})_{\a}^{\; \; \b}
= - i \ve^{ijk} (\s^k)_{\a}^{\; \; \b}
\ee
\be
(\s^{i})_{\a}^{\; \; \b}
(\s^{i})_{\g}^{\; \; \d}
= 2 \d^{\d}_{\a} \d^{\b}_{\g}
-  \d^{\b}_{\a} \d^{\d}_{\g}
\ee

Let us use the following shorthand:
\be
A \cdot \s^{\m} \cdot \ov{B}
=
A^{\a} \s^{\m}_{\a \dot{\b}} \ov{B}^{\dot \b}
, \;
\ov{A} \cdot {\ov \s}^{\m} \cdot B
=
\ov{A}^{\dot \a} {\ov \s}^{\m }_{ \dot \a \b} B^{ \b}
\ee
The following identities help to familiarize the notation:
\be
A \cdot B = - B \cdot A = A^{\a} B_{\a}
\ee
\be
\ov{A} \cdot \ov{B} = - \ov{B} \cdot \ov{A} =  \ov{A}^{\a} \ov{B}_{\a}
\ee
\be
\c \cdot \y =  \y \cdot \c
= \y^{\a} \c_{\a}
\ee
\be
\ov{\c} \cdot \ov{\y} =  \ov{\y} \cdot \ov{\c}
=
\ov{\y}^{\dot \a} \ov{\c}_{\dot \a}
\ee
\be
(\s^{\m} \cdot \ov{\s}^{\n} \cdot \s^{\l})_{\a \dot \d}
=
(\s^{\m})_{\a \dot \b} (\ov{\s}^{\n})^{\dot \b \g} (\s^{\l})_{\g \dot \d}
\ee
where we use latin letters for commuting spinors and greek letters for
anticommuting ones.
It should be remembered that since $\ve_{\a \b}$ is antisymmetric,so
\be
\c_{\a} \y^{\a} = - \c^{\a} \y_{\a}
\ee
Formulae involving products of these invariant tensors can
be reduced using the basic relations:
\be
\s^{\m} \cdot \ov{\s}^{\n} \cdot \s^{\l}
+
\s^{\l} \cdot \ov{\s}^{\n} \cdot \s^{\m}
=
2 \eta^{\m \l} \s^{\n}
- 2 \eta^{\m \n} \s^{\l}
-
2 \eta^{\l \n} \s^{\m}
\ee
\be
\s^{\m} \cdot \ov{\s}^{\n} \cdot \s^{\l}
-
\s^{\l} \cdot \ov{\s}^{\n} \cdot \s^{\m}
=
- 2i \ve^{\m \n \l \r} \s_{\r}
\ee
where we define:
\be
\ve^{0ijk} =
\ve^{ijk}
\ee
Similarly one gets:
\be
(A \cdot \s^{\m} \cdot \ov{B} )^*
=
B \cdot \s^{\m} \cdot \ov{A}
=
\ov{A} \cdot \ov{\s}^{\m} \cdot B
\ee
and, in particular,
\be
(A \cdot \s^{\m} \cdot \ov{A} )^*
=
A \cdot \s^{\m} \cdot \ov{A}
=
\ov{A} \cdot \ov{\s}^{\m} \cdot A
\ee
is a real quantity.
The Fierz identity takes the form:
\be
A^{\a} \s^{\m}_{\a \dot \b}
 \ov{B}^{\dot \b}
C^{\g} \s_{\m \g \dot \d} \ov{D}^{\d}
=
-
2 A^{\a} C_{\a} \ov{B}^{\dot \b} \ov{D}_{\dot \b}
\ee
or
\be
A \cdot \s^{\m}\cdot
\ov{B}
C \cdot  \s_{\m}
\cdot \ov{D}
=
-
2 A\cdot  C \; \ov{B}
\cdot  \ov{D}
\ee
for commuting spinors, with appropriate change of sign for
the anticommuting case.

\section{Inner Product}
\setcounter{equation}{0}
\renewcommand{\theequation}{B.\arabic{equation}}

Here we discuss the inner product used in the text.
It is at first surprising that a positive
metric in the Fock space can be defined while preserving the
non-compact Lorentz invariant metric.
 The reason this can be done is that the
two metrics are not connected in any way. One is a Fock space metric defined
for arbitrary polynomials, and the other is a restriction on the space
of polynomials.  Here are some examples.

  We define the adjoint spinor $(c^{\a})^{\dag}$ to satisfy:
\be
[ (c^{\a})^{\dag},
c^{\b} ]
=
\d^{\b}_{\a}
\ee
\be
[ (\ov{c}^{\dot \a})^{\dag},
\ov{c}^{\dot \b} ]
=
\d^{\dot \b}_{\dot \a}
\ee
\be
[ (\ov{c}^{\dot \a})^{\dag},
c^{\b} ]
=
0
\ee
Now consider for example the following expression:
\[
\langle 0 |
( c^{\a} \s^{\m}_{\a \dot \b} \ov{\l}^{\dot \b})^{\dag}
( c^{\g} \s^{\n}_{\g\dot \d} \ov{\l}^{\dot \d})
| 0
\rangle
\]
\be
=
(\s^{\m}_{\a \dot \b})^{\dag} \s^{\n}_{\g\dot \d}
\d^{\g}_{\a}
\d^{\dot \d}_{\dot \b}
=
(- \ov{\s}_{\m}^{\dot \b \a })
\s^{\n}_{\a \dot \b}
= + 2 \d^{\n}_{\m}
\ee
The equation
\be
(\s^{\m}_{\a \dot \b})^{*}
=
(- \ov{\s}_{\m}^{\dot \b \a })
\ee
is a consequence of the hermiticity of the $\s^{\m}$ matrices
and of the identities (\ref{-sigma}) and (\ref{sigma}).
Here * means the complex conjugation operator needed
when taking the hermitian conjugate of an operator in
Fock space. Writing the indices in this way
enables one to see that the positive definite
inner product preserves the Lorentz invariance
of Lorentz invariant expressions.

Another example is the following. The
Minkowski metric controls the way that indices are contracted
and this is preserved by the operators.  The Fock space metric
assigns a positive number to each vector.  Take for example the state
\be
< p_{\m} p^{\m} |
 p_{\n} p^{\n} >
=
<0| (p^{\m})^{\dag}
(p_{\m})^{\dag}
p_{\n} p^{\n}|0 >
= 2
\h_{\m \n}
\h^{\m \n}
= 8
\ee
where we use
\be
[ p_{\m}^{\dag} , p_{\n}] = \d^{\m}_{\n}
\ee

\end{document}